\documentclass[conference,a4paper]{APSIPA2026}
\usepackage{amsmath}
\usepackage{graphicx}
\usepackage{booktabs}
\usepackage{multirow}
\usepackage{threeparttable}
\usepackage{cite}
\usepackage{url}

\usepackage{geometry}
\usepackage{fancyhdr}

\fancypagestyle{firststyle}{
  \fancyhf{}
}

\begin{document}

\title{Contextual Biasing for Streaming ASR via CTC-based Word Spotting}

\author{
\authorblockN{Kai-Chen Tsai, Tien-Hong Lo, Yun-Ting Sun, and Berlin Chen}
\authorblockA{
Department of Computer Science and Information Engineering,
National Taiwan Normal University\\
\{41247065s, teinhonglo, 61347015s, berlin\}@ntnu.edu.tw
}
}

\maketitle
\thispagestyle{firststyle}
\pagestyle{empty}

\begin{abstract}
Contextual biasing is essential to improving the recognition of rare and domain-specific words in an automatic speech recognition (ASR) system. While numerous methods have been proposed in recent years, most of them focus on offline settings and do not explicitly address the challenges of streaming ASR. For example, CTC-based word spotting (CTC-WS) have demonstrated strong performance by directly detecting keywords from CTC log-probabilities, but they are limited to offline processing and require access to the full utterance. In This work, we present a streaming extension of CTC-WS for real-time contextual biasing. Our method maintains active keyword paths across audio chunks using a stateful token passing algorithm, enabling the detection of keywords that span multiple chunks. To ensure low latency and stable output, we introduce an incremental commitment mechanism that only emits segments guaranteed not to be affected by future audio, while deferring uncertain regions. This method naturally integrates with streaming ASR pipelines and does not require modifications to the underlying acoustic model or additional training, making it practical for real-world deployment. Experimental results show that our method reduces overall WER and effectively improves keyword F-score, demonstrating its effectiveness for real-time ASR applications.
\end{abstract}

\begin{IEEEkeywords}
Contextual biasing, Streaming ASR, CTC, RNN-T, Key-word Spotting. 
\end{IEEEkeywords} 

\section{introduction}
While automatic speech recognition (ASR) models have improved significantly in recent years, they still face challenges in accurately recognizing biasing words, including context-specific and rare words such as person names and place names\cite{ner_survey}. Incorrect recognition can cause problems in real-world applications, including captioning and keyword-based interaction scenarios, where accurate recognition of predefined terms is critical. Therefore, contextual biasing, which leverages a predefined list of terms to improve their recognition accuracy, has become increasingly important in ASR systems.

In addition to recognition accuracy, streaming ASR is also a major topic in recent years. Most ASR are designed for full audio input, which is impractical for lots of scenarios such as live captioning, voice assistants, online meetings, and human-computer interaction, where recognition results must be produced incrementally with low latency. In such streaming settings, the system cannot wait until the entire utterance is available before performing recognition. Instead, it must process audio chunk by chunk while maintaining stable and accurate outputs. Therefore, how to combine streaming ASR and contextual biasing well remains a challenging and practical problem. 

\begin{figure}[t]
    \centering
    \includegraphics[width=\columnwidth]{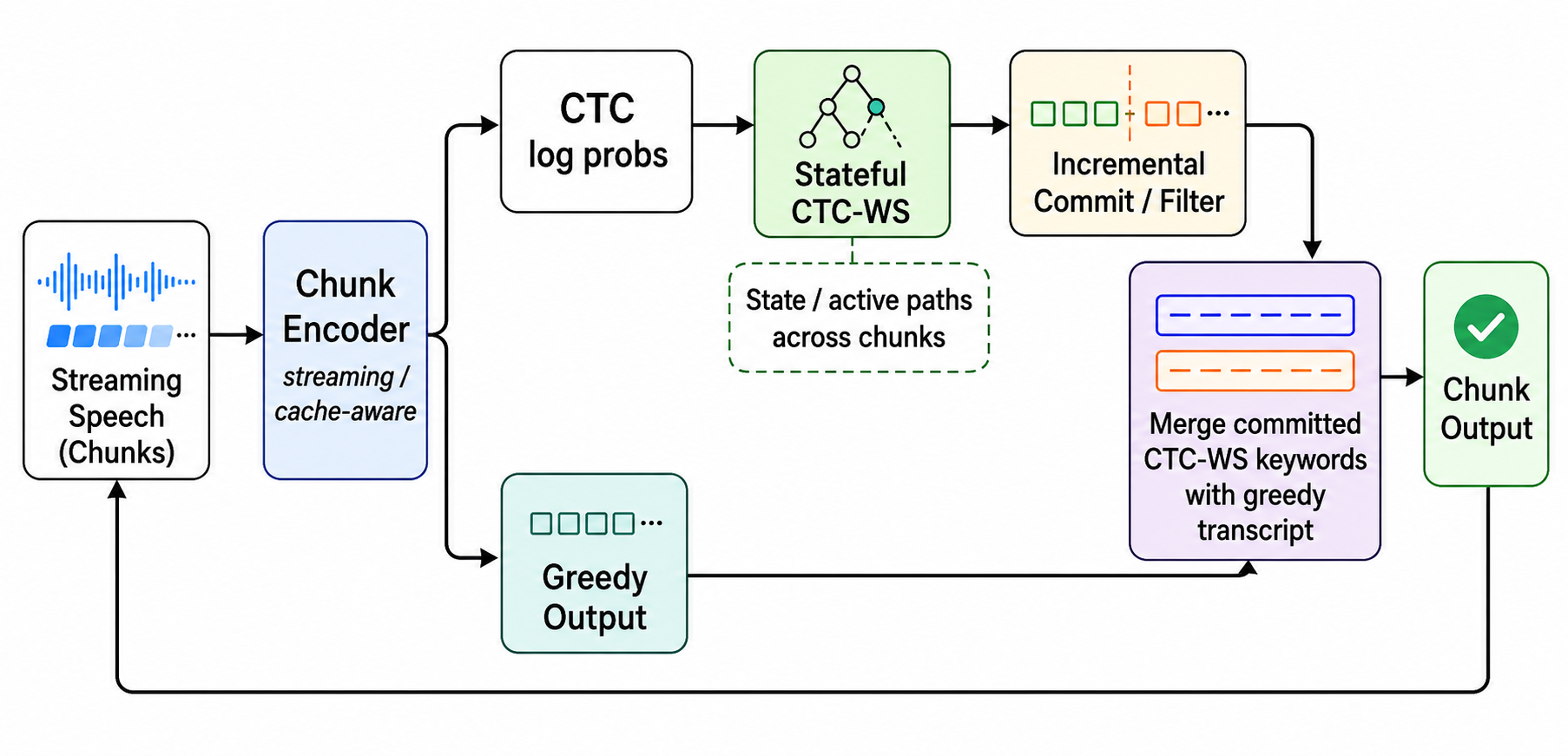}
    \caption{The proposed Streaming CTC-WS pipeline.}
    \label{fig:pipeline}
\end{figure}

Various contextual biasing methods have been proposed, and they can be categorized according to how contextual information is incorporated into the recognition process. 

One representative direction is deep-fusion-based contextual biasing, which introduces biasing information directly into the neural ASR model. In these methods, the biasing list is encoded as additional contextual representations and injected into the encoder or decoder through mechanisms such as cross-attention~\cite{b1,b2,b3}, or through an additional context module trained for contextual biasing~\cite{b4,b5,b6}. Since contextual information is integrated inside the model, these methods usually require additional model components, model-specific integration, or changes to the training process. Although such methods can be applied to streaming ASR, the need for architectural modification or additional training makes them less flexible when the ASR model is treated as a fixed recognizer.

Another common direction is shallow-fusion, where biasing information is incorporated during decoding step. In this type of approach, an external language model or biasing graph is used to adjust the decoding score of hypotheses that contain predefined terms~\cite{b7,b8,b9,b10}. Shallow fusion usually does not require changing the internal architecture of the ASR model, making it easier to apply to an already trained recognizer. However, since it typically relies on beam search to effectively boost target phrases, applying it in a streaming setting introduces additional latency. While shallow fusion can also be applied in greedy decoding mode to avoid this issue, the absence of multiple hypotheses reduces its ability to accurately recognize target phrases.

Beyond fusion-based methods, CTC-based Word Spotting (CTC-WS)\cite{b11} provides another direction for Connectionist Temporal Classification (CTC)\cite{b12} and Recurrent Neural Transducer (RNN-T)\cite{rnnt} model. Instead of incorporating biasing information through model fusion or beam-search rescoring, CTC-WS uses CTC log probabilities and trie to detect predefined terms from the biasing list. This makes it attractive for CTC-based ASR systems because it can be applied without retraining the acoustic model or substantially modifying the ASR architecture. However, the original CTC-WS method assumes access to the complete audio sequence, making its direct application to streaming ASR non-trivial.

To address this limitation, this work extends CTC-based Word Spotting to the streaming ASR setting (Figure 1). The key challenge is that streaming ASR processes speech incrementally, so a biasing word may be split across multiple chunks and cannot always be detected from a single chunk alone. Therefore, the proposed method maintains active word-spotting states across chunks, allowing partial matches to continue when new audio becomes available. In addition, an incremental commitment mechanism is introduced to decide which parts of the recognition result are stable enough to be emitted, while keeping uncertain regions pending for future chunks. In this way, the proposed method enables CTC-WS-based contextual biasing for streaming ASR without retraining the acoustic model or changing the basic ASR recognition framework.

\section{CTC-WS}
CTC-WS is a contextual biasing method that operates on the CTC log probabilities produced by an CTC model. Given a predefined biasing list, CTC-WS constructs a trie-based context graph to represent the token sequences of target words or phrases. The graph is combined with the CTC transition topology, allowing the search process to handle blank symbols and repeated tokens in CTC outputs.

During word spotting, the CTC log probabilities are searched frame by frame along the context graph. Active hypotheses are extended according to the token probabilities and valid graph transitions. When a hypothesis reaches the end of a biasing word or phrase, it is recorded as a candidate with an accumulated score and a corresponding frame interval. To encourage the detection of biasing words, a context-biasing weight can be added when hypotheses move through non-blank tokens.

After candidate generation, overlapping detections are filtered by keeping the best-scoring candidate. The remaining candidates are then compared with the word-level alignment from greedy ASR decoding. If a candidate is more reliable than the overlapping greedy recognition result, it can replace the original word in the final transcription. This comparison helps reduce false positives caused by acoustically similar words.

The main advantage of CTC-WS is that it avoids beam-search rescoring and directly detects predefined terms from CTC probabilities. Therefore, it can recover biasing words that may not appear in the greedy decoding output while keeping the decoding process efficient. However, the original CTC-WS method assumes that the complete CTC log-probability sequence is available before word spotting. This makes it suitable for offline recognition, but difficult to apply directly to streaming ASR, where audio is processed chunk by chunk and biasing words may span across chunk boundaries.

\begin{figure}[t]
    \centering
    \includegraphics[width=\columnwidth]{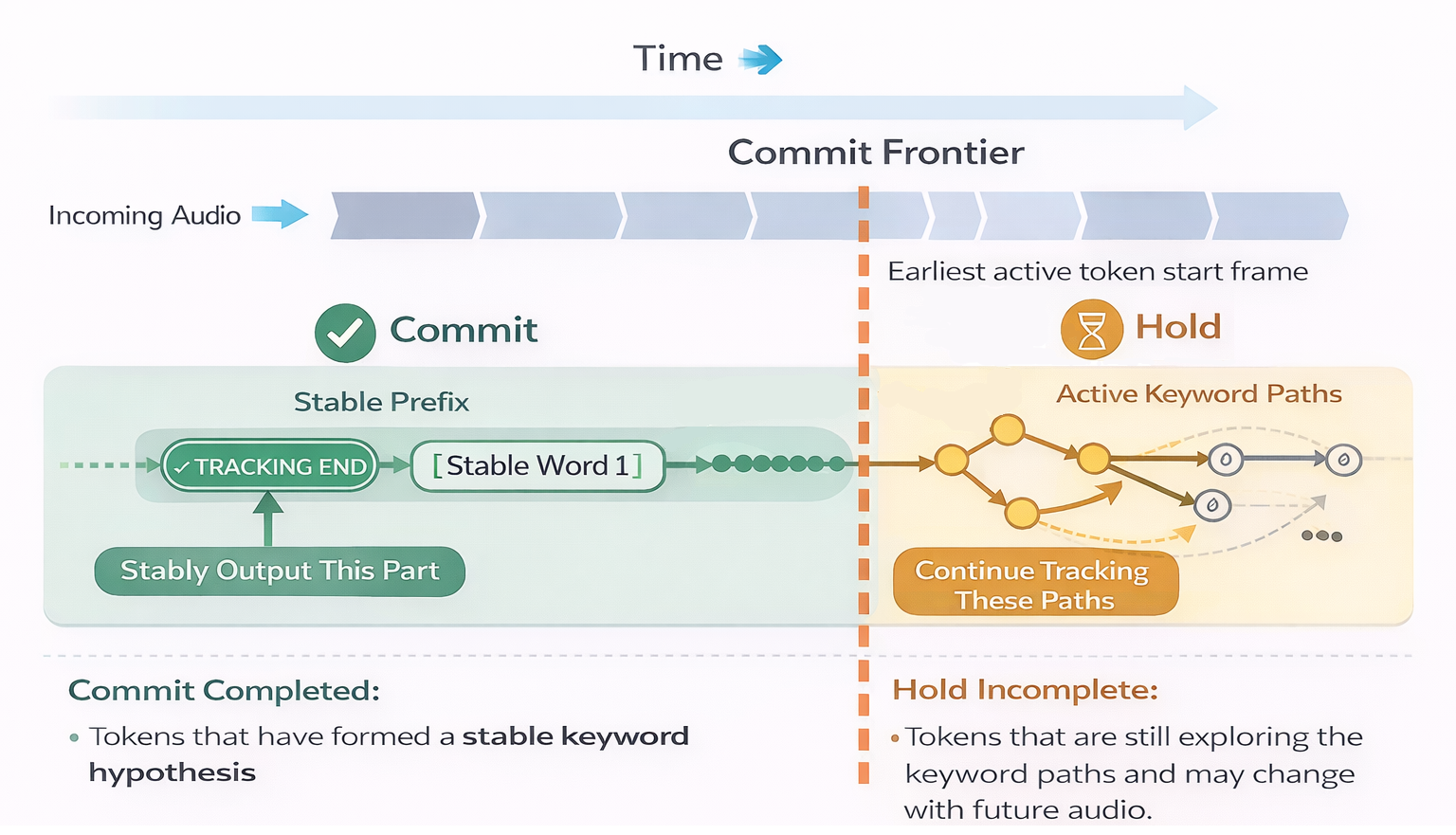}
    \caption{Commit and Hold for Cross-Chunk Keyword Tracking--- Completed keyword hypotheses before the commit frontier are safely emitted or used to replace the greedy ASR output. Active paths after the frontier are held and continued with future chunks until they become safe to commit.}
    \label{fig:detail}
\end{figure}

\section{integration with streaming ASR}
To address this issue, we propose a streaming extension of CTC-WS that preserves the original CTC-based spotting procedure while changing the state management and output commitment strategy. The proposed method maintains active word-spotting hypotheses across chunks so that incomplete matches near chunk boundaries are not discarded. It further introduces an incremental commitment mechanism (Figure \ref{fig:detail}), which emits only the part of the recognition result that is no longer affected by active hypotheses.

\subsection{Stateful Word Spotting across Chunks}
In offline CTC-WS, the token passing algorithm is applied to the complete CTC log-probability sequence. In the streaming setting, the input audio is processed chunk by chunk, so directly applying the offline algorithm to each chunk independently may discard unfinished keyword matches at chunk boundaries.

To address this issue, the proposed method maintains active tokens across chunks. At each chunk, the streaming ASR model produces CTC log probabilities, and the word spotter updates both newly initialized tokens and active tokens inherited from previous chunks. Each active token records its current state in the context graph, accumulated score, and start frame. When an active token reaches the end of a biasing word, it is stored as a spotted keyword candidate with its score and frame interval. If it remains incomplete at the end of the chunk, it is preserved and passed to the next chunk.

This design keeps the core token passing algorithm the same as offline CTC-WS, while allowing keywords that span multiple chunks to be detected.

\subsection{Incremental Commitment Mechanism}
Since some active tokens may still lead to future keyword detections, the proposed method does not immediately finalize all spotted candidates after each chunk. Instead, it divides the timeline into a committed region and a hold region. The committed region contains frames that are considered safe to output, while the hold region contains recent frames that may still be affected by unfinished keyword matches.

The commit frontier is defined according to the earliest start frame among the remaining active tokens. Frames before this frontier cannot be affected by any future keyword completion, because all active tokens that could still produce future candidates start at or after the frontier. Therefore, candidates located before the commit frontier can be safely finalized. If no active token remains, the commit frontier is moved to the current frame offset.

This mechanism allows the system to output stable segments incrementally while holding only the uncertain region near the current processing boundary.

\subsection{Online Keyword Replacement}
After the commit frontier is determined, the proposed method performs keyword replacement only within the committed region. Spotted candidates whose frame intervals are fully inside the committed region are selected first. If multiple candidates overlap, only the best-scoring candidate is retained by the streaming word spotter, following the same de-overlap principle as offline CTC-WS.

The selected candidates are then merged with the decoder word alignment in the committed region.  Given a spotted candidate and an aligned decoder word, the replacement decision is based on their frame-level intersection percentage. A word is replaced only when the candidate interval covers more than a predefined percentage of the word interval. In our implementation, this threshold is set to 50\%.

The merged result is emitted as the committed streaming output, while the hold region is kept pending for future chunks. At the end of the utterance, all remaining active tokens are cleared, the hold region is flushed, and the final biased transcription is produced.

\section{experiments setup}
\subsection{Streaming ASR Model}
In this work, we used the publicly available NVIDIA Streaming model\footnote{\url{https://huggingface.co/nvidia/stt_en_fastconformer_hybrid_large_streaming_multi}},which uses a FastConformer encoder\cite{b13} with a hybrid CTC / RNNT decoding architecture and supports cache-aware streaming inference\cite{b14} with around 114M parameters and is trained for streaming ASR. The model was trained on approximately 20k hours of English speech data, using a BPE tokenizer\cite{b15} with 1024 tokens. It supports multiple look-ahead configurations, which allows the model to operate under different latency settings. The cache-aware streaming mechanism constrains the available past and future contexts in the encoder and reuses cached activations from previous chunks during inference. In this way, the model can process speech chunk by chunk without recomputing the overlapping context, making streaming inference more efficient.

\begin{table}[t]
    \centering
    \caption{Statistics of the evaluation datasets.}
    \label{tab:dataset_statistics}
    \begin{tabular}{lcc}
        \toprule
        \textbf{Metric} & \textbf{STOP1} & \textbf{STOP2} \\
        \midrule
        Duration (h) & 3.15 & 7.21 \\
        Utterances & 3413 & 7384 \\
        Bias list size & 687 & 1107 \\
        \midrule
        Example phrases 
        & \begin{tabular}{@{}c@{}}
            halsey \\
            justin bieber \\
            the doobie brothers
          \end{tabular}
        & \begin{tabular}{@{}c@{}}
            africa \\
            batikkumeli street \\
            southern united states
          \end{tabular} \\
        \bottomrule
    \end{tabular}
\end{table}

\subsection{Datasets}
To evaluate the recognition performance of specific words, we used two datasets that contain a large number of named entities.
\begin{itemize}
    \item \textbf{STOP1}\footnote{ \url{https://github.com/GLCLAP/GLCLAP-stop1-stop2-dataset}}: This dataset is derived from the STOP\cite{b16} test and evaluation sets, with filtering applied to retain only queries containing person names.
    \item \textbf{STOP2}: This dataset is constructed in the same way as STOP1, but the filtering criterion is changed to location-name entities. 
\end{itemize}

The biasing list for each dataset is built from the target entities appearing in that dataset. Specifically, STOP1 uses person names appearing in its utterances as biasing words, while STOP2 uses location names appearing in its utterances.

\subsection{Metrics}
For biasing-word evaluation, we measure the recognition accuracy of biasing words using the F-score (\(2 \times \mathrm{Precision} \times \mathrm{Recall} / (\mathrm{Precision} + \mathrm{Recall}) \times 100\%\)), over the key phrases in the context-biasing list.
We also computed the overall WER to evaluate the general recognition quality of the ASR output. In addition, we measure the per-chunk computation time of each processing step. All evaluations were run on Intel Core i7-8700 and NVIDIA GeForce RTX 2080 Ti.

\section{Results}

\begin{table*}[t]
\centering
\begin{threeparttable}
\caption{Recognition performance on STOP1 and STOP2, and the chunk size is set to 1120 milliseconds. All values are reported in percentages. P and R denote Precision and Recall, respectively.}
\label{tab:stop-results}

{
\normalsize
\setlength{\tabcolsep}{13pt}
\begin{tabular}{ll|c@{\hspace{10pt}}c|c@{\hspace{10pt}}c}
\toprule
\textbf{Decoder} & \textbf{Method}
& \multicolumn{2}{c|}{\textbf{STOP1}}
& \multicolumn{2}{c}{\textbf{STOP2}} \\
\cmidrule(lr){3-4} \cmidrule(lr){5-6}
& 
& \textbf{WER$\downarrow$} & \textbf{F-score (P/R)$\uparrow$}
& \textbf{WER$\downarrow$} & \textbf{F-score (P/R)$\uparrow$} \\
\midrule
\multirow{3}{*}{CTC}
& ---        & 18.36 & 66.84 (96.87/51.02) & 12.09 & 88.26 (98.76/79.79) \\
& GPU-PB           & 15.42 & 78.78 (96.39/66.62) & 11.86 & 90.53 (98.05/84.09) \\
& Streaming CTC-WS & \textbf{12.83} & \textbf{89.61} (93.87/85.72) 
                   & \textbf{10.48} & \textbf{95.06} (97.35/92.88) \\
\midrule
\multirow{3}{*}{RNN-T}
& ---        & 15.46 & 74.08 (96.89/59.96) & 10.43 & 92.16 (98.60/86.51) \\
& GPU-PB           & 12.21 & 84.88 (95.91/76.12) & 10.18 & 93.62 (96.83/90.62) \\
& Streaming CTC-WS & \textbf{12.09} & \textbf{88.67} (91.51/86.00) 
                   & \textbf{9.66} & \textbf{95.60} (97.47/93.79) \\
\bottomrule
\end{tabular}
}

\end{threeparttable}
\end{table*}

\subsection{Effectiveness of Contextual Biasing}
Table~\ref{tab:stop-results} shows the recognition performance with CTC and RNN-T decoding on STOP1 and STOP2. For both decoding settings, Streaming CTC-WS achieves the best results across the two datasets. With CTC decoding, it reduces WER from 18.36\% to 12.83\% and improves the F-score from 66.84\% to 89.61\% on STOP1. On STOP2, it reduces WER from 12.09\% to 10.48\% and improves the F-score from 88.26\% to 95.06\%. With RNN-T decoding, the same trend is observed, where Streaming CTC-WS also obtains the lowest WER and highest F-score on both datasets. The improvement mainly comes from higher recall, indicating that the proposed method can recover more biasing words from the context-biasing list.

We further compare our method with GPU-PB~\cite{b10}, a GPU-accelerated phrase-boosting method. Similar to shallow fusion, GPU-PB applies contextual biasing during decoding by adding scores from a phrase-boosting tree. Unlike conventional shallow fusion methods that often rely on beam search and may slow down decoding, GPU-PB uses a GPU-accelerated boosting tree with a modified score distribution, allowing phrase boosting to be used efficiently even in greedy decoding. In our streaming experiments, GPU-PB is applied in greedy mode to satisfy the streaming setting.

GPU-PB also improves over the non-biasing baseline, showing that phrase boosting is useful for recognizing predefined entities in streaming ASR. However, Streaming CTC-WS consistently achieves lower WER and higher F-score than GPU-PB under both CTC and RNN-T decoding. This difference mainly comes from the biasing mechanism. GPU-PB modifies token scores during greedy decoding, but greedy decoding keeps only one active hypothesis at each step. Therefore, if the target phrase is not selected during decoding, phrase boosting may have limited ability to recover it. In contrast, Streaming CTC-WS directly detects biasing words from CTC log probabilities and merges the detected candidates with the greedy output, allowing it to recover more biasing words missed by the original streaming result.
\begin{figure}[t]
    \centering
    \includegraphics[width=\columnwidth]{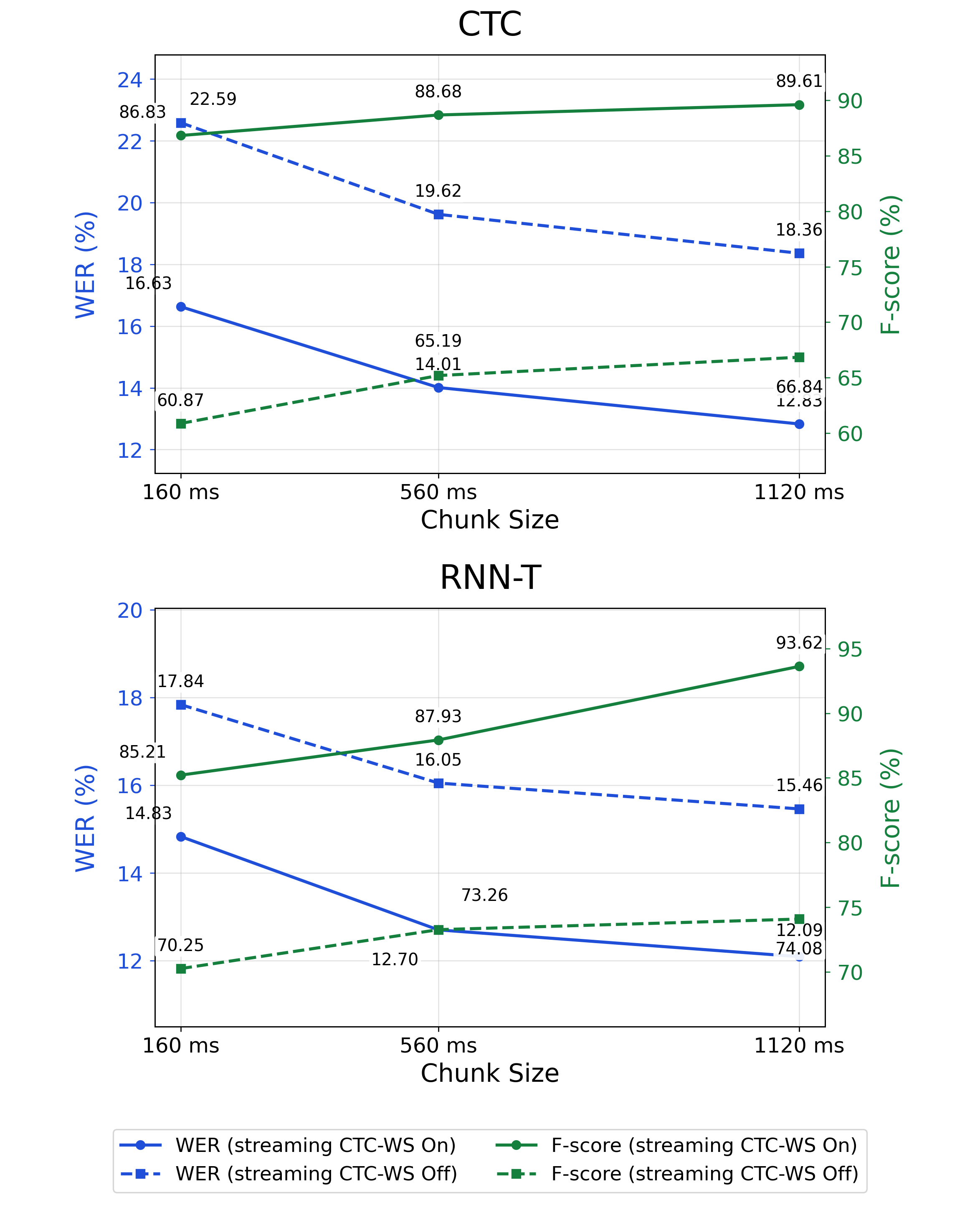}
    \caption{Effect of chunk size on WER and F-score with and without contextual biasing test on STOP1. CB means if the proposed method is used.}
    \label{fig:chunk-size}
\end{figure}
\subsection{Evaluation on different chunk size}
To analyze the effect of streaming chunk size, we evaluate three chunk-size settings: 160 ms, 560 ms, and 1120 ms.

Figure~\ref{fig:chunk-size} shows the effect of chunk size on WER and F-score with and without contextual biasing under CTC and RNN-T decoding. Compared with the non-biasing baseline, Streaming CTC-WS consistently achieves lower WER and higher F-score across all chunk sizes for both decoding settings. The improvement is especially clear in F-score. Without contextual biasing, the F-score drops significantly under smaller chunk sizes, indicating that short chunks make biasing-word recognition more difficult. In contrast, Streaming CTC-WS maintains much higher F-scores even at 160 ms, showing that the proposed stateful word-spotting mechanism can preserve active keyword paths across chunks and recover biasing words that may be split by chunk boundaries.

The best performance is obtained with the 1120 ms chunk size for both CTC and RNN-T decoding. In this setting, Streaming CTC-WS achieves the lowest WER and highest F-score. Meanwhile, the results also show that the proposed method remains effective under smaller chunk sizes, suggesting that it can improve biasing-word recognition across different streaming latency settings.

\subsection{Runtime Analysis}
Table~\ref{tab:runtime} shows the per-chunk runtime analysis under different chunk sizes. The reported time includes both ASR computation and the additional time introduced by the proposed method. Compared with CTC inference, the proposed method adds computation mainly from word spotting, while merging remains negligible. Although this increases the total processing time over CTC-only inference, the extra cost is small relative to the chunk duration. We also measured the runtime with RNN-T decoding and observed a similar values, indicating that the added overhead mainly comes from the proposed word-spotting and merging steps rather than the decoder choice.

Specifically, the mean extra processing time accounts for only 4.1\%, 3.6\%, and 3.2\% of the chunk duration for 160 ms, 560 ms, and 1120 ms chunks, respectively. Even at P95, the extra processing time remains below 9\% of the chunk duration across all settings. These results indicate that the proposed Streaming CTC-WS introduces limited streaming overhead and remains practical for real-time inference.  
\begin{table}[t]
\centering
\caption{Runtime analysis of streaming CTC-WS under different chunk sizes. Each value is reported as mean / P95. The extra processing ratio is computed as the additional processing time beyond CTC inference divided by the chunk duration.}
\label{tab:runtime}
\resizebox{\columnwidth}{!}{
\begin{tabular}{lccc}
\toprule
\textbf{Metric} & \textbf{160 ms} & \textbf{560 ms} & \textbf{1120 ms} \\
\midrule
CTC Time (ms) & 38.3 / 41.3 & 41.6 / 44.8 & 45.9 / 47.3 \\
Spot Time (ms) & 5.2 / 11.4 & 18.8 / 39.6 & 34.3 / 72.2 \\
Merge Time (ms) & 0.8 / 1.2 & 0.9 / 1.3 & 1.0 / 1.7 \\
Total Time (ms) & 44.9 / 55.2 & 61.9 / 80.6 & 81.9 / 119.9 \\
\midrule
Extra Proc. / Chunk (\%) & 4.1 / 8.7 & 3.6 / 6.4 & 3.2 / 6.5 \\
\bottomrule
\end{tabular}
}
\end{table}

\section{Conclusion}
This work extends CTC-based Word Spotting to streaming ASR for contextual biasing. By maintaining active word-spotting states across chunks, the proposed method can detect biasing words that span chunk boundaries. An incremental commitment mechanism is also introduced to emit stable regions while holding incomplete keyword paths for future chunks. Experimental results show that the proposed Streaming CTC-WS improves biasing-word recognition while maintaining practical streaming efficiency. Overall, this method provides a simple and effective approach for integrating contextual biasing into streaming ASR.

\end{document}